\documentclass[twocolumn,showpacs,prl,aps]{revtex4}
\usepackage[T1]{fontenc}
\usepackage[latin1]{inputenc}
\usepackage{epsfig,graphics}

\providecommand{\LyX}{L\kern-.1667em\lower.25em\hbox{Y}\kern-.125emX\@}
\newcommand{\noun}[1]{\textsc{#1}}

\begin{document}

\title{Dual coherent particle emission as generalised Cherenkov-like
effect\\ in high
energy particle collisions }

\author{D. B. Ion} 
  \email{Dumitru.Ion@cern.ch}
  \affiliation{National Institute for Physics and Nuclear Engineering,
 NIPNE-HH,
 Bucharest, P.O. Box MG-6, Romania\\ and \\
TH Division, CERN, CH-1211 Geneva 23,
Switzerland
}

\author{E. K. Sarkisyan}
  \email{sedward@mail.cern.ch}
  \affiliation{EP Division, CERN, CH-1211 Geneva 23, 
Switzerland\\ and\\ High Energy Physics Institute Nijmegen (HEFIN),
University of Nijmegen/NIKHEF, NL-6525 ED Nijmegen, The Netherlands}

\date{\today}%

\begin{abstract}

In this paper we introduce a new kind of nuclear/hadronic coherent
particle production mechanism in high-energy collisions called \emph{dual
coherent particle emission (DCPE)} which takes place when the phase
velocities of the emitted particle $v_{Mph}$ and that of
particle source $v_{B_1ph}$  satisfy the dual coherence condition:
$v_{Mph}\leq v_{{B_1}ph}^{-1}$. The general signatures of the
DCPE in the nuclear and hadronic media are
established and some experimental evidences are given.

\end{abstract}


\maketitle

Cherenkov radiation was first observed in the early 1900's by the experiments
developed by Mary and Pierre Curie when studying radioactivity emission. The
nature of such radiation was unknown until the experimental works (1934-1937)
of P.A. Cherenkov and the theoretical interpretation by I. E. Tamm and I. M.
Frank (1937) when it was clarified that this radiation is produced by
charged
particles travelling through matter at speeds larger than the phase velocity
of light in the medium. 
Although it was not well understood and recognised
until 1958 (Nobel Prize in Physics),
the effect afterwards was called the Cherenkov radiation (CR).
The idea that meson production in nuclear interaction may be
described as a process similar to the electromagnetic CR
go back to Wada, Ivanenko and Gurgenidze, and Blohintzev and
Indenbom
(see Refs. {[}1{]}) who propagated it first on a qualitative basis and after
that by using some nuclear models. In the last decade this idea of
the generalisation
of the CR to mesons and photons inside the nuclear and hadronic media was
systematically
investigated  theoretically {[}2-4{]}  and also tested
experimentally
{[}5-8{]}.
Such idea was also extended to a Cherenkov-like mechanism for meson
production in
hadron-hadron interactions {[}8-11{]} at high energies. In this analogy, a fast
hadron is the source of mesonic field as it is the fast charged particle
in the
normal Cherenkov radiation. Thus the mesonic Cherenkov-like effect might be
expected to occur when a {\it hadron traverses the nuclear (or hadronic)
matter with
a velocity exceeding the phase velocity of the mesonic field}. 

In this paper, we introduce a new kind of coherent particle production
mechanism
in particle collisions called \emph{dual coherent particle emission
(DCPE)}
from which all kind of the generalised Cherenkov-like effects can be
obtained.
The DCPE effects are expected to take place when the phase
velocities
of the emitted particle \( v_{Mph} \) and that of particle source
\( v_{B_1ph} \)
satisfy the {\it dual coherence condition}: \( v_{B_{1}ph}v_{Mph}\leq 1\). 
We investigate the
general signatures of DCPE effects in nuclear and hadronic media and a
propose their experimental tests. The
comparison
of the absolute predictions {[}3{]} on the nuclear pionic Cherenkov-like
effect
(NPICR) with the recent experimental results obtained in relativistic
nuclear collisions {[}5-7{]}
is given. 

\vspace*{.2cm}

\emph{\noun{Dual coherent particle emission}} \noun{(DCPE)}. Let us start
with a general B\( _{1}\rightarrow {\rm MB} \)\( _{2} \) decay described
in 
Fig. \ref{fig:1}(a). Here
 a particle M {[}with energy \( \omega , \) momentum $k=$ Re\(
n_{M}(\omega ) \)\( \sqrt{\omega ^2-M^2_{\rm M}}\),
rest mass \(M_{\rm M} \), and refractive index \( n_{M}(\omega )] \) is
emitted
in a (nuclear, hadronic, dielectric, etc.) medium from an incident particle
B\( _{1} \) {[}with energy \(E_1 \), momentum \(p
_{1}={\rm Re} n(E_{1})\sqrt{E^{2}_{1}-M_{1}^{2}} \),
rest mass M\( _{1} \), and refractive index \( n_{1}(E_{1}) \){]} that
itself
goes over into a final particle B\( _{2} \) {[}with energy \( E_{2} \),
momentum
\(p _2={\rm Re}n(E_{2})\sqrt{E^{2}_{2}-M_{2}^{2}} \),  rest
mass 
\(M_{2} \),
and refractive index \(n_{2}(E_{2}) \){]}.

{\par
\begin{figure}
\centering \resizebox*{4.1cm}{6.4cm}{\includegraphics{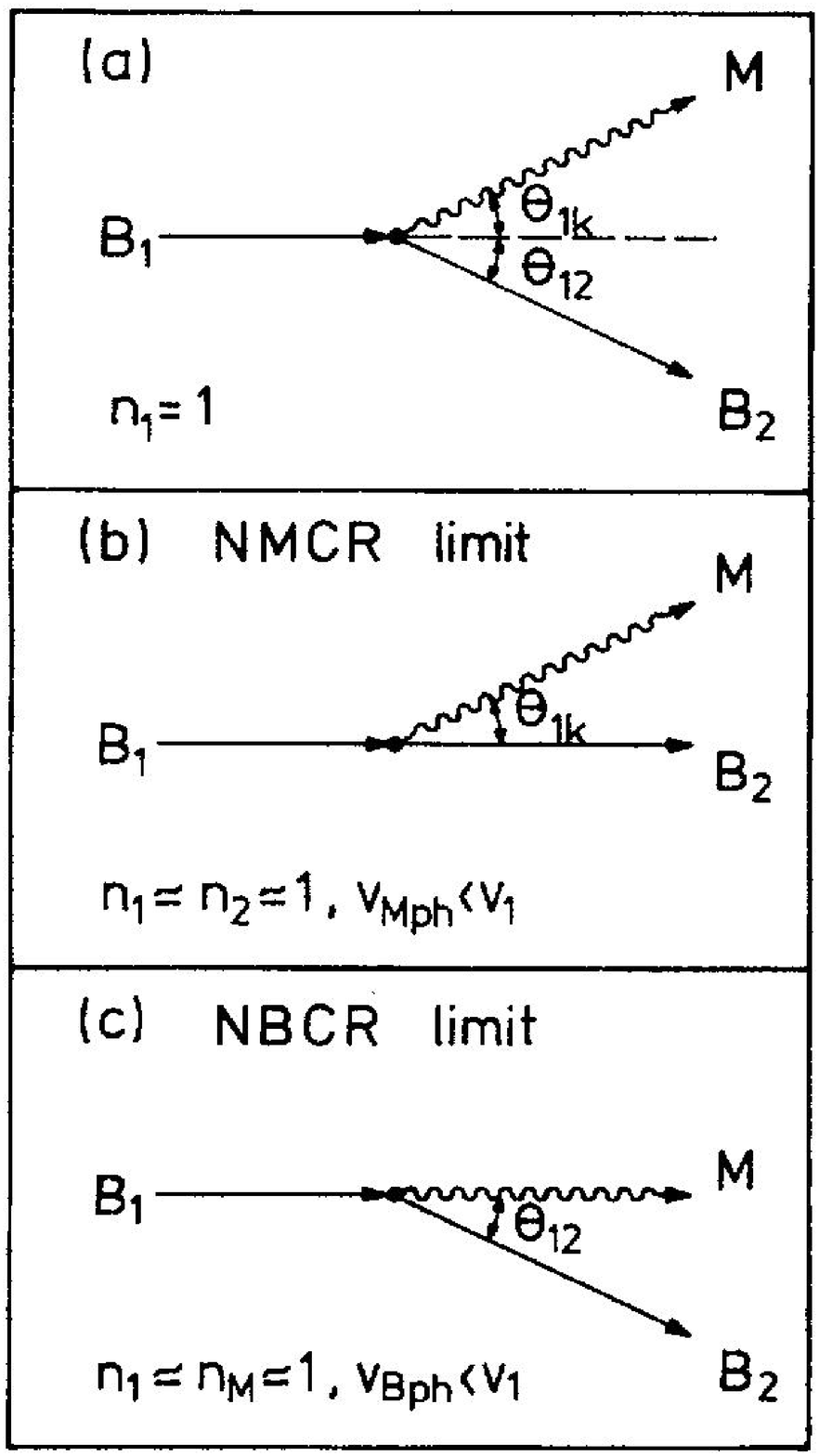}  }
\caption{(a) The two-body decay process B$_{1}$ $\rightarrow {\rm MB}_2$
in
 nuclear
media. (b) The nuclear mesonic Cherenkov-like radiation (NMCR) limit
{[}3{]}:
\( v_{Mph}\leq v_{1}. \) (c) The nuclear baryonic Cherenkov-like
radiation
limit {[}3{]}: \( v_{B_{2}ph}\leq v_{1} \)
}
\label{fig:1}
\end{figure}
\par}

Here we prove that in order to obtain a genuine spontaneous particle emission
in a given medium {\it the two general conditions} are necessary to be
fulfilled: 

\begin{itemize}
\item {(i) The incident particle-source must be coupled to a specific radiation
field (RF)} (see Fig. \ref{fig:1}) and that {the particles
propagation properties
in medium must be modified.}
\item {(ii.1) The particle source must be moving in medium with a dual phase
velocity \( v^{-1}_{B_{1}ph} \) higher than the phase velocity \(v_{Mph}
\)
of the RF-quanta.}
\item {(ii.2) The particle source must be moving in medium with a dual phase
velocity \( v^{-1}_{B_{1}ph} \) higher than \( v_{B_{2}ph} \).}
\end{itemize}
\emph{Proof:} The propagation properties of particles in a medium are changed
in agreement with their elastic scattering with the constituents of that medium.
So, the phase velocity \(v_{Xph}(E_X) \) of any particle X 
(with the
total
energy $E_X$ and rest mass \(M_{X} \)) 
in medium is
modified
as follows:\begin{equation}
\label{Eqo}
v_{Xph}(E_{X})=\frac{E_{X}}{p_{X}}=\frac{1}{{\rm 
Re}n_{X}(E_{X})}.\frac{E_{X}}{[E_{X}^{2}-M^{2}_{X}]^{\frac{1}{2}}}
\end{equation}
The \emph{refractive index} \( n_{X}(E_X) \) 
in a medium composed from the
constituents
``c'' can be calculated in standard way by using the Foldy-Lax formula \(
[12
\){]}
(we work in the units system \( {\hbar}=c=1 \)) \begin{equation}
\label{Eq0}
n^{2}_{X}(E_{X})=1+\frac{4\pi \rho
}{E_{X}^{2}-M^{2}_{X}}\cdot C(E_{X})\overline{f}_{Xc\rightarrow Xc}(E_{X})
\end{equation}
 where \( \rho  \) is the density of the constituents, $C(\omega)$
is
a \emph{coherence factor} [$C(\omega )=1$ when the the medium
constituents
are randomly distributed], \( \overline{f}_{Xc\rightarrow Xc}(E_X) \) is
the
averaged
forward Xc-scattering amplitude. 

Now, by using the \emph{energy-momentum conservation law} for the decay
\( {\rm B}_1 \rightarrow {\rm MB}_{2} \)
in medium,
\begin{equation}
\label{10}
E_{1}=E_{2}+\omega ,\; \; \overrightarrow{p}_{1}=\overrightarrow{p}_{2}+\overrightarrow{k}
\end{equation}
 we obtain (see angles definition in Fig. \ref{fig:1}):
\begin{equation}
\label{Eq10a}
cos\theta
_{1k}=v_{Mph}v_{B_{1}ph}+\frac{1}{2p_{1}k}[-D_{B_{1}}+D_{B_{2}}-D_{M}]
\end{equation}
\begin{equation}
\label{Eq10c}
cos\theta _{12}=v_{B_{1}ph}v_{B_{2}ph}+\frac{1}{2p_{1}k}[-D_{B_{1}}-D_{B_{2}}+D_{M}]
\end{equation}
\begin{equation}
\label{Eq10b}
cos\theta _{2k}=v_{Mph}v_{B_{2}ph}+\frac{1}{2p_{2}k}[D_{B_{1}}-D_{B_{2}}-D_{M}]
\end{equation}
 where \(D_{X} \),\( \; X\equiv B_{1},B_{2},M, \) are departures from mass
shell, and are given by

\begin{equation}
\label{Eq1c}
D_{X}\equiv
E^{2}_{X}-p^{2}_{X}=M^{2}_{X}+[1-({\rm 
Re}n_{X}(E_{X}))^{2}][E_{X}^{2}-M^{2}_{X}]
\end{equation}

We note that the second terms in the right side of Eqs (4)-(6) can be considered
as quantum corrections to the first {}``classical{}'' terms [3]. 

Now, a rigorous proof of the statement (ii) is obtained from conditions that
the respective emission angles must be the physical angles. The
\emph{coherence
quantum conditions:} \( cos \theta _{ij} \) \( \leq 1, \) $i,j=1,2,k,$
(4)-(6),
at the high incident particles energies transform into \emph{classical
coherence conditions},
e. g., \begin{equation}
\label{Eq1d}
cos\theta _{1k}=v_{Mph}v_{B_{1}ph}\leq 1
\end{equation}
 which is equivalently to

\begin{equation}
\label{Eq1e}
v_{Mph}\leq v^{-1}_{B_{1}ph},\; {\rm or}\; \; v_{B_{1}ph}\leq v^{-1}_{Mph}
\end{equation}

It is worth to note that from the \emph{dual coherence conditions}
(9) we obtain the following DCPE condition.

\begin{itemize}
\item {(iii) In case when \( v^{-1}_{B_{1}ph} =v _{1} \),
from
the condition (ii) the  two important generalised Cherenkov-like limits
follow:
(iii.1) the NMCR Cherenkov-like radiation limit \( v_{Mph}  \leq v_{1}
\), and
(iii.2) the NBCR Cherenkov-like radiation limit \( v_{B_{2}ph}\leq v_{1}
\).}
\end{itemize}
The proof of the statement (iii) is obtained immediately if one observes
that when
the particle \emph{B\( _{1} \)} is \emph{on the mass shell in medium
{\rm (Re}n\(
_{1}=1) \)
then \( v^{-1}_{B_{1}ph} \)=v\( _{1} \),} and the \emph{dual coherence conditions}
(ii.1) and (ii.2) will go into the \emph{Cherenkov-like coherence conditions}
(iii.1) and (iii.2), respectively. 

For each kind of generalised Cherenkov-like effect there is a \emph{threshold
projectile kinetic energy} given by 
\begin{equation}
\label{Eq5}
T_{thr}(E_{X})=M_{p}\sqrt{1-[v_{Xph}(E_{X})]^{2}}-1].
\end{equation}

Now, we can obtain a classification of these DCPE effects not only on the
basis of four fundamental (strong, electromagnetic, weak,
and gravitational)  interactions but also using the above
\emph{``M-B duality''} as well as the \emph{crossing
symmetry.}

The main signatures of the (DCPE) as generalised Cherenkov-like effects are
as follows:

\begin{itemize}
\item The  differential cross sections posses the bumps in the energy
bands where the
DCPE-coherence conditions are fulfilled.
\item The DCPE-effects are threshold mechanisms.
\item Coherent particles emitted via the DCPE mechanism must be coplanar
with
the incoming
and outgoing projectiles: strong \( (\theta _{1k},\omega ) \) and
\( (\theta _{1k},E_{p})\) correlations.
\item The intensities as well as the absorption effects can be calculated  
as in the case
of generalised Cherenkov-like effects {[}3{]} by using Feynman diagrams in
medium. 
\item Any two-body decay process B$_1$ $\rightarrow$ MB$_2$  in
medium
via DCPE
mechanism posses two limiting modes: the NMCR mode in which the particle M
is
emitted
when the coherence condition \( v_{Mph}\leq v_{1} \)is fulfilled, and
the NBCR mode
when the coherence condition \( v_{B_{2}ph}\leq v_{1}\) is fulfilled.
\end{itemize}
\noun{Coherent meson emission in nuclear media}. The lightest mesons for which
the elementary scattering amplitudes on nucleons in the forward direction are
well known from the experimental data are pions {[}13{]}. In the last
decade,
the \emph{nuclear pionic Cherenkov-like radiation (NPICR)} was intensively investigated
theoretically {[}2,3{]} on the basis of elementary pion-nucleon scattering
amplitudes
in the forward direction. The characteristic features of the NPICR-pions predicted
in Refs. {[}2,3{]} are as follows:

\begin{itemize}
\item {The numerical values on the refractive index of pions in nuclear
medium was
obtained in the standard way by applying Lax-Foldy formula (2) with the coherence
factor $C=1$.}
\item {The NPICR-coherence condition} \( v_{ph}(\omega )\leq v \) ,
 was
found to be {\it fulfilled in the three energy bands:}\\ {\hspace*{.1cm} 
(CB1) 190 MeV\(\leq \omega \leq \) 315 MeV, for all \( \pi ^{\pm ,0} \);}\\ 
{\hspace*{.1cm} (CB2) 910 MeV \( \leq \omega \leq  \) 960 MeV, only for \(
\pi \)\( ^{+}
\);}\\ 
{\hspace*{.1cm} (CB3) \( \omega  \geq  \)
80 GeV, for all \( \pi ^{\pm ,0} \).}\\
{The values for the NPICR-thresholds for all the three pionic
bands are
shown in Fig. \ref{fig:2}.}
\item {The NPICR-pions must be coplanar with the incoming and outgoing
projectile
possessing strong \( (\theta _{1k},\omega ) \) and \( (\theta
_{1k},T_{p})\)
 correlations 
(see Fig. 7 from Ref. [3]).}
\item The \emph{NPICR-differential cross sections (DCS)} are peaked
at
the energy \( \omega _{m}=244 \) MeV for CB1-emission band when absorption
in medium is taken into account. The CB1-peak width in the DCS is predicted
to be \( \Gamma _{m}  \leq  \) 25 MeV (see details in Figs. 8 and 9
from Ref.{[}3{]}).
\item The energy behaviour of NPICR-peak position is
predicted to be
as \( T^{-2}_{p} \) (see Fig. 12 in {[}3{]}), while the $A$-target
dependence of the NPICR differential cross sections is given in Fig. 14
of [3]. 
\end{itemize}
The existence of the NPICR-CB1-emission band {[}2,3{]} was recently confirmed
experimentally by the  JINR-Dubna group {[}5,6{]} in the studies of 
 Mg-Mg central collisions at 4.3 GeV/$c$ per nucleon by processing the
pictures
from the
2m Streamer Chamber SKM-200. After processing a total number of 14218
events,
which were found to meet the  centrality criterion,  the
experimental
results shown in Fig.3 have been obtained.

{\par
\centering 
\begin{figure}
\epsfig{file=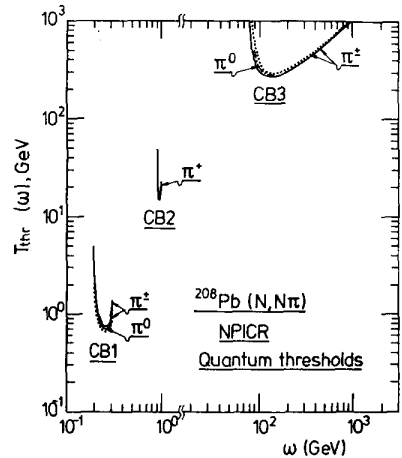,
width=6.cm,height=7.cm,
clip=} 
\caption{
The absolute predictions for the threshold T\(
_{thr}(\omega ) \) for
single coherent meson production via \emph{pionic Cherenkov-like
radiation (NPICR)}
mechanism {[}3{]} in the nuclear reaction \( ^{208}\)Pb 
(p$\rightarrow$
N\(\pi  \)).}
\label{fig:2}
\end{figure}
\par}

\par{}

\par

{\par
\centering 
\begin{figure}
\resizebox*{0.47\textwidth}{0.3\textheight}
{\includegraphics{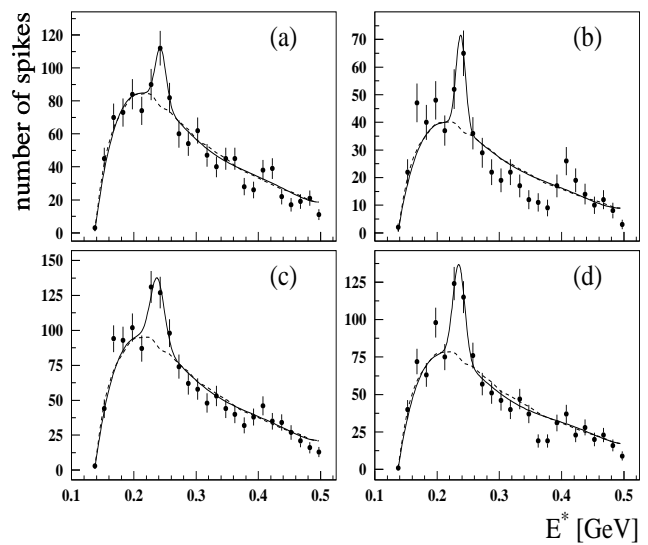} 
}
\caption{The energy distributions obtained {[}5,6{]} for the \( \pi ^{-}
\)s produced
in the central Mg-Mg collisions at 4.3 GeV/$c$ per nucleon.}
\label{fig:3}
\end{figure}
\par}

\par{}

\par{} \vspace{0.3cm}

As one sees from Fig. 3, the energy distributions of emitted pions in the
central collisions have a significant peak over the inclusive background.
The
experimental values of the peak position and its width are 
\begin{equation}
\label{Eq7a}
E^{*}_{m}=238\pm 3({\rm stat})\pm 8({\rm syst})\: {\rm MeV},
\end{equation}
\begin{equation}
\label{Eq.7b}
\Gamma _{m}=10\pm 3({\rm stat})\pm 5({\rm syst}) \: {\rm MeV}.
\end{equation}
 Hence, the \emph{values of the peak energy and its width} obtained in
Refs.
{[}5,6{]} {\it are in a good agreement with the absolute NPICR 
predictions for $\omega_m and \Gamma_m$}
{[}2,3{]}.
It is important to note that the above value of \( E^{*}_{m} \) is similar
to the position of the peak observed in Ref. {[}15{]} in the analysis
of \( \pi ^{+} \)production in coincidence measurements of (p,n) reactions
at 0\( ^{0} \) on C, the effect connected with the NPICR mechanism {[}3{]}.

Therefore, we conclude that the NPICR cross sections in all emission
bands, as well as other NPICR-signatures, are large enough in order
to be experimentally measured in exclusive experiments. Of the
great importance is the experimental discovery of the first coherent NPICR
band in the inclusive measurements since it  represents the first
experimental proof that the generalised Cherenkov-like effect is the
real mechanism for the coherent particles production with the rest mass
different from zero.

\vspace{.3cm}

\emph{\noun{Generalised mesonic Cherenkov-like radiation in hadronic media.}}
The mesonic Cerenkov-like radiation in hadronic media was considered by many
authors {[}8-11{]}. A systematic investigation of the classical and
quantum
theory of this kind of effects in hadronic media can be found in Ref.
{[}9{]}. The classical variant {[}2,9{]} of the Cherenkov mechanism was applied
to the study of single meson production in hadron-hadron interactions at high
energies. This variant is based on the usual assumption that hadrons are composed
from a central core in which most of the hadron mass is concentrated surrounded
by a large and more diffuse mesonic cloud (hadronic medium). Then, 
it was shown [9,11]
that a \emph{hadronic mesonic Cherenkov-like radiation (HPICR)} with an
mesonic
refractive index given by a pole approximation, can be able to explain
the integrated
cross sections of a single meson production in hadron-hadron interaction.
As an example, in Fig. \ref{fig:4} we present the integrated cross
sections of the process
\( \rm pp\rightarrow pp\pi ^{0} \), compared with the HPICR-predictions in
the
\emph{limit of the HPICR mechanism dominance}. 
This result was very encouraging for the extension of HPICR-dominance
hypothesis to other single meson production in hadron-hadron
collisions at high energies. Collecting all the \( \chi ^{2}/dof \)  for
all 139 reaction fitted with
the \emph{hadronic mesonic Cherenkov-like} (HMCR) mechanism, we get the
surprisingly good description as shown in  Fig.  \ref{fig:5}. We must
underline that only reactions with single  meson
production was fitted (a single parameter fit) with the HMCR predictions
on the integrated cross sections. 

{\par
\begin{figure}
\centering 
\resizebox*{5.6cm}{5.6cm}{\includegraphics{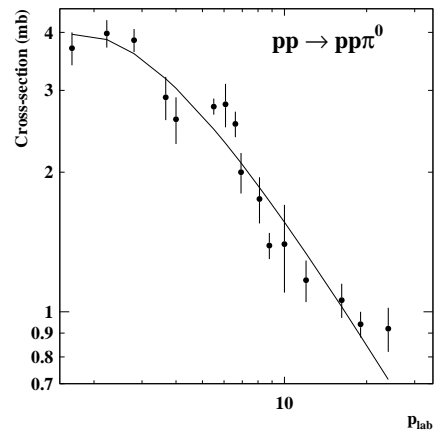}  } 
\caption{Experimental statistical test of the
\emph{HMCR}-\emph{mechanism dominance}
in hadron-hadron collisions.}
\label{fig:4}
\end{figure}
\par}

Also, recently, in hadronic experiments [5-7,16], the Cherenkov-like
radiation in the variant
proposed in \cite{10}, has been observed. 

{\par
\begin{figure}
\centering
\resizebox*{5.6cm}{5.6cm}{\includegraphics{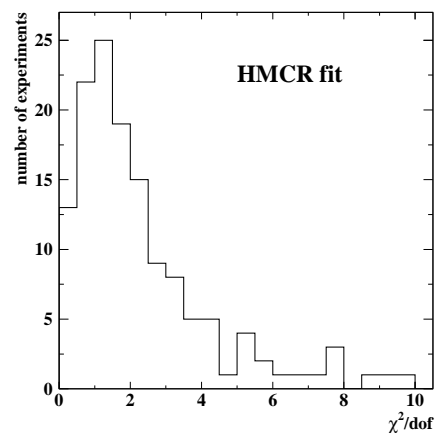} 
}
\caption{
The number of reactions fitted with
HMCR-predictions
for integrated cross sections as a function of \( \chi ^{2}/dof
\).}
\label{fig:5}
\end{figure}
\par}

\par{} \vspace{0.3cm}

\noun{Conclusions.} In this paper a new mechanism called \emph{dual coherent
particle emission} (DCPE) is introduced. \noun{}

The results and conclusion obtained in this paper are as follows:

\begin{itemize}

\item The DCPE mechanism  includes in
a more general and exact way all
the types of the Cherenkov mechanisms thanks to the \emph{general dual
coherence condition} \( v_{B_{1}ph}v_{Mph}\leq 1 \) {which has} the
\emph{particular limiting cases:}  \emph{M-Cherenkov-like radiation
limit: v\( _{Mph} \)\( \leq v_{1} \), and}  \emph{B-Cherenkov-like
radiation limit: v\( _{B_{2}ph}\leq v_{1} \).}

\item In the \emph{nuclear media} we presented the first experimental
results {[}5,6{]} which confirm with high accuracy the predictions {[}3{]}
in the first CB1-pionic band.

\item In the \emph{hadronic media} we presented the \emph{first
statistical test} (see Figs. \ref{fig:4} and \ref{fig:5}) based on the
fit of the
integrated cross sections of 139 reaction of single meson production.
These results suggest a possible \emph{HMCR-dominance} of the
single meson production in high energy hadron-hadron collision. 

\end{itemize}

Finally, it is important to note that special experimental techniques
based on the coincidence measurements are necessary in order to extract
the
yields of the DCPE generalised Cherenkov-like effects from the background
produced by other mechanisms.

\begin{acknowledgments}
We would like to thank G. Altarelli for fruitfull discussions. One of the
authors (D.B.I.) would like to thank TH Division for hospitality during
his stay at CERN. One of the authors (E.K.S.) would like to thank the
"Nederlandse 
Organisatie voor Wetenschappelijk Onderzoek (NWO)" for support.
\end{acknowledgments}

\end{document}